\newif\iftwocolumn
\newcommand{\figwidth}{\iftwocolumn \columnwidth \else 100mm\fi}

\documentclass[aip,rsi,amsmath,amssymb,reprint,groupedaddress]{revtex4-1}\twocolumntrue

\usepackage{graphicx}
\usepackage{placeins}
\usepackage{amsmath,amssymb}
\usepackage[colorlinks=true,citecolor=red]{hyperref}		% adds hyperlinks to .pdf file
\usepackage{natbib}
\usepackage{xcolor}
\usepackage{siunitx}

\definecolor{groen}{RGB}{0,150,0}
\newcommand{\chk}[1]{{\color{blue}#1}}

\begin{document}
\title{A digital feedback controller for stabilizing large electric currents to the ppm level for Feshbach resonance studies}

\author{R. Thomas}\email{ryan.j.thomas1@gmail.com}
\author{N. Kj{\ae}rgaard}\email{nk@otago.ac.nz}
\affiliation{Department of Physics, QSO-Centre for Quantum Science, and Dodd-Walls Centre, University of Otago, Dunedin, New Zealand}

\begin{abstract}
Magnetic Feshbach resonances are a key tool in the field of ultracold quantum gases, but their full exploitation requires the generation of large, stable magnetic fields up to 1000 G with fractional stabilities of better than $10^{-4}$.  Design considerations for electromagnets producing these fields, such as optical access and fast dynamical response, mean that electric currents in excess of 100 A are often needed to obtain the requisite field strengths.  We describe a simple digital proportional-integral-derivative current controller constructed using a field-programmable gate array and off-the-shelf evaluation boards which allows for gain scheduling, enabling optimal control of current sources with non-linear actuators.  Our controller can stabilize an electric current of 337.5 A to the level of $7.5\times 10^{-7}$ in an averaging time of 10 minutes and with a control bandwidth of 2 kHz.  

\end{abstract}

\maketitle
\section{Introduction}
\label{sec:Introduction}
Feshbach resonances are ubiquitous in the field of ultracold atomic physics\cite{FeshbachReview} as they allow the interaction strength between atoms to be changed\cite{Inouye1998}.  This tunability can be used to explore effects such as the BEC-BCS phase transition in degenerate Fermi gases\cite{Greiner2003,Bartenstein2004,Bourdel2004}, Anderson localization\cite{Roati2008}, and Efimov trimers\cite{Kraemer2006,Knoop2009}.  Sweeping a magnetic field across a Feshbach resonance is used to produce and study ultracold molecules\cite{Thompson2005,Regal2003} which have complicated scattering properties\cite{DeMarco2019,Gregory2019,Christianen2019}.  Feshbach resonances have also been critical in the study of dipolar quantum gases\cite{Ferrier-Barbut2016,Lahaye2008,Koch2007}.  Finally, Feshbach resonances can be used as a sensitive probe of variations in fundamental constants such as the ratio of electron to proton mass\cite{Chin2006}.

Magnetic Feshbach resonances in alkali metal systems are typically accessed by using a pair of Helmholtz coils to generate stable fields ranging up to 1000 G.  The required field stability, measured by fluctuations $\delta B$, is determined by two factors.  The first is that $\delta B \ll |\Delta|$ where $\Delta$ is the width of the resonance which can range from 10 mG to 100 G\cite{FeshbachReview}.  This level of stability is easily achieved for broad Feshbach resonances, but is more difficult for narrow resonances and especially resonances in non-zero angular momentum channels\cite{Cui2017,Yao2019} or in alkaline earth atoms\cite{Barbe2018}.  The second factor is related to how one measures the magnetic field \emph{in situ}; typically, the field is measured by probing the transition frequency between two magnetic sub-levels of trapped atoms using either Rabi or Ramsey spectroscopy.  For both of these techniques, fluctuations in  the transition frequency $\delta f$ at a magnetic field $B_0$ corresponding to the Feshbach resonance need to satisfy $\delta f = \left.\frac{df}{dB}\right|_{B=B_0}\delta B \lessapprox \Omega/(2\pi)$ where $\Omega$ is the Rabi frequency of the transition which is limited by the particular transition and available microwave or radio-frequency power.  As a concrete example, consider the relatively broad Feshbach resonance between $^{87}$Rb and $^{40}$K atoms\cite{DeMarco2019} located at $B_0=546$ G with width $\Delta=-3$ G and suppose a Rabi frequency of $\Omega =2\pi\times 5$ kHz.  If we require $\delta B \approx \Delta/20$, then the fractional stability $\delta B/B_0$ needed is about 270 ppm.  However, calibration of the field at $546$ G using radio-frequency spectroscopy of the least magnetically-sensitive transition in $^{40}$K with $df/dB \approx 0.1$ MHz/G implies a needed magnetic field stability better than 100 ppm, so for this example the stability criterion for calibrating the field is stricter than the stability needed for using the resonance.  As an alternate case, consider the relatively narrow Feshbach resonance between two $^{87}$Rb atoms located at 1007 G with a width of 210 mG\cite{Volz2003,Durr2004}.  Here, having $\delta B \approx \Delta/20$ implies a fractional stability of 10 ppm which is stricter than the 17 ppm stability needed for measurement using the least-sensitive transition.
  
In an experiment, one typically stabilizes the electric current generating the magnetic field and not the magnetic field itself.  Although a number of superb current sources in the 10 mA to 10 A range have been demonstrated\cite{Fan2019,Wang2015,Erickson2008}, design considerations for Helmholtz coils in ultracold atomic physics experiments, such as optical access to atomic samples, fast dynamic response, and adequate heat removal mean that these coils are often large with few windings which implies that currents in the hundreds of amperes are needed to produce magnetic fields in the 100 G -- 1000 G range\cite{Chin2013,Zhaoyuan2014,Roux2019}.  Therefore, the task of stabilizing a large magnetic field to the level of 10 ppm is essentially equivalent to that of stabilizing a large ($> 100$ A) current to the same level, although further reductions in magnetic field noise can be achieved by measuring and stabilising the magnetic field directly\cite{Merkel2019}.  A typical solution is to build an analog proportional-integral-derivative (PID) feedback loop that stabilizes the current produced from a voltage-controlled power supply using external regulating transistors to control the current.  While some of these solutions have been very successful, with reported stabilities of better than 10 ppm\cite{Marte2003,Trenkwalder2011,Zhaoyuan2014,Wacker2015,Shuai2019,Yu-Meng2019}, designing and implementing a high-precision, low-noise, and robust analog PID controller from discrete, linear components is a non-trivial task because the transistors used to regulate the current are inherently non-linear.  A better solution in these situations is to use so-called ``gain scheduling'' where the PID gain parameters are changed depending on the state of the system according to a linearised description of the system dynamics\cite{Bechhoefer2005,Rugh2000}.  Gain scheduling is a standard technique in control theory, and it has been used in applications ranging from inductive cooking appliances\cite{Jimenez2014} to magnetic levitation\cite{Bojan-Dragos2016}.

In this article, we present a digital PID controller that uses off-the-shelf components to regulate the electric current from a relatively noisy (100 ppm) commercial power supply.  We achieve real-time performance by using a field-programmable gate array (FPGA) to measure the current, generate a dynamic control signal, vary the loop parameters, and implement the PID algorithm.  Gain scheduling is a key feature of our FPGA controller which makes it ideally suited for managing a non-linear system.  Our system can stabilize a current of 337.5 A to the level of approximately 250 \si{\micro\ampere} in an averaging time of 10 minutes with a control bandwidth of 2 kHz, corresponding to a fractional magnetic field stability of $0.75$ ppm.  

\section{Implementation}
\label{sec:Implementation}
\begin{figure*}[t]
\includegraphics[width=\textwidth]{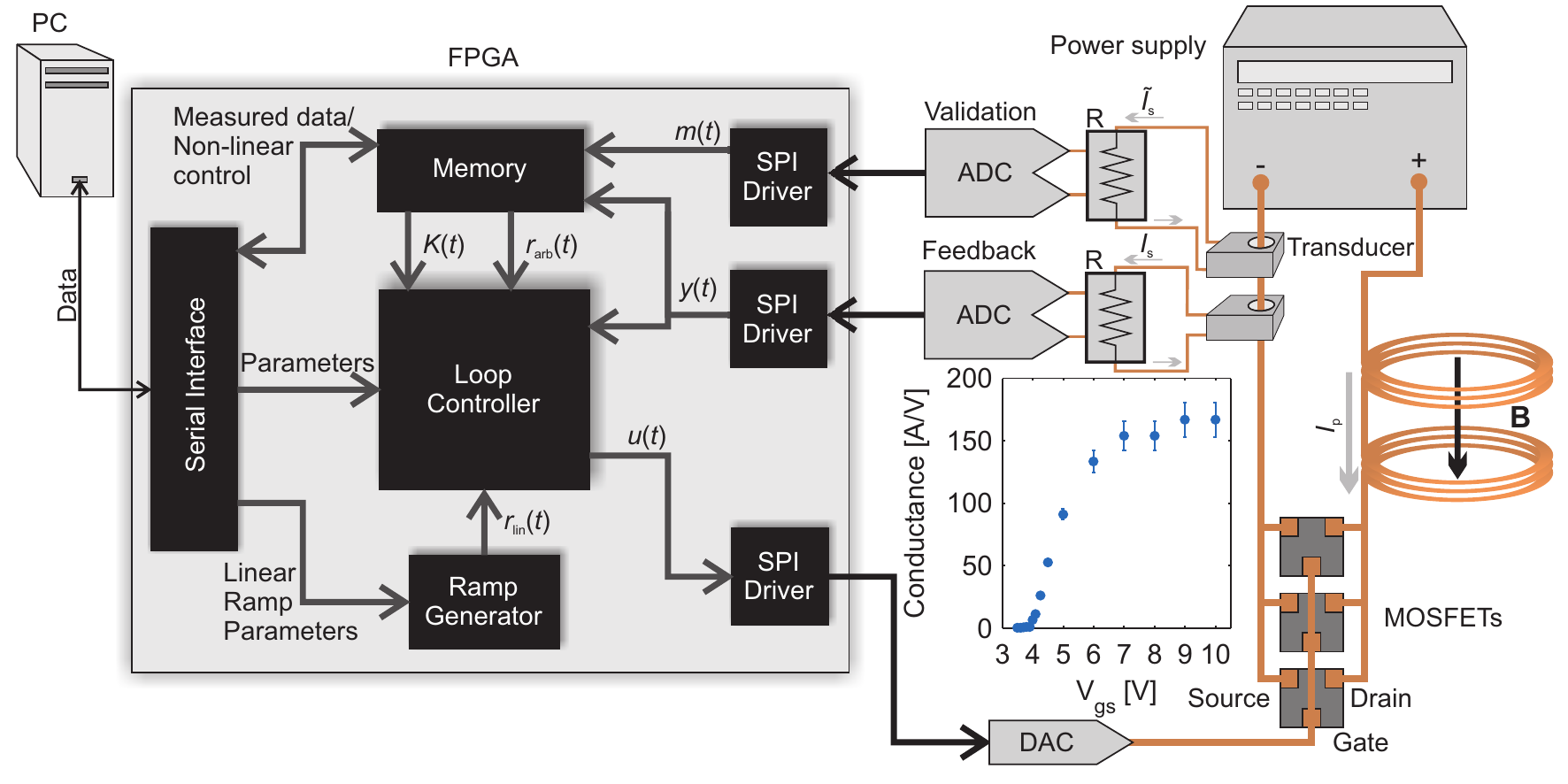}
\caption{Architecture of the PID controller as implemented in controllable logic and a diagram of the physical connections.  The primary current $I_p$ through a pair of Helmholtz coils is measured by using a fluxgate current transducer to produce a secondary current $I_s$ that generates a voltage drop across a four-terminal precision resistor that is then digitized by an ADC.  This feedback measurement $y(t)$ is compared to either a linear control voltage $r_{\rm lin}(t)$ or an arbitrary control voltage $r_{\rm arb}(t)$, and an actuator value $u(t)$ is produced that is sent to a DAC.  The loop controller uses either fixed loop parameters or a time-varying sequence of parameters $K(t)$ from memory.  The voltage output of the DAC drives the gates of three high-current MOSFETs that control the flow of $I_p$.  An additional transducer, resistor, and ADC produces a signal $m(t)$ which we use for validating our servo's performance.  A serial connection with a PC is used to set parameters and control voltages and to retrieve ADC measurements.  Inset plot shows the measured drain-source conductance for a single MOSFET as a function of gate-source voltage.}
\label{fg:Diagram}
\end{figure*}

Single-input single-output control loops can be broken into four parts: a control value/set-point $r(t)$ for the state of the system, a measurement $y(t)$ of that state, an actuator signal $u(t)$ that affects the state, and a control law $K$ that relates $u(t)$, $r(t)$, and $y(t)$\cite{Bechhoefer2005}.  At the heart of our control loop is a Xilinx Spartan 6 FPGA mounted on a commercial development board (Numato Labs Saturn) with 512 Mb of LPDDR RAM.  As shown schematically in Fig.~\ref{fg:Diagram}, the FPGA implements both the control law and the control signal in programmable logic, handles communication with the measurement and actuator electronics, stores measurement and control values in memory, and interfaces with a PC over a USB cable using the universal asynchronous receiver/transmitter (UART) protocol\cite{GitHub}.

The physical system to be controlled is shown on the right-hand side of Fig.~\ref{fg:Diagram}.  A 15 V/440 A power supply (Agilent 6690A) sources the so-called ``primary'' current $I_p$ that passes through a pair of Helmholtz coils and produces a magnetic field $\mathbf{B}$ at the atoms in our experiment.  The supply is set to voltage-controlled mode, and the value of $I_p$ is controlled on the `low-side'' of the Helmholtz coils using three n-channel MOSFETs (IXYS IXFN200N07) connected in parallel with 25 mm x 5 mm copper bus bars.  We connect transient voltage suppressing diodes (STMicroelectronics BZW50-47B) across the drain and source of each MOSFET to suppress voltage spikes when the MOSFETs are turned off.  The gate-source voltage of these transistors serves as the actuator signal $u(t)$ and changes the effective drain-source conductance which changes the primary current (inset of Fig.~\ref{fg:Diagram}).  The gate voltage is sourced by a high-precision, 20-bit digital-to-analog converter (DAC) mounted on a manufacturer-supplied evaluation board (Analog Devices EVAL-AD5791SDZ), and communication with the DAC uses the serial peripheral interface (SPI).  The board is populated with nearly all the necessary components to produce a voltage in the range of $-10$ V to $10$ V, and we use an external voltage reference (Maxim Integrated MAX6350) to supply the DAC with its 5 V reference.

We measure the primary current using a flux gate current transducer (Danisense IDSA600) which generates a secondary current $I_s = I_p/1500$ that passes through a precision 10 $\Omega$ sense resistor (Vishay Y169010R0000T9L, $0.01\%$ tolerance, $0.2$ ppm/K) in the four-terminal Kelvin configuration to produce a sense voltage.  We digitize this voltage to produce the feedback measurement $y(t)$ using a low-noise, 24-bit, differential analog-to-digital converter (ADC) mounted on an evaluation board (Texas Instruments ADS127L01EVM).  The evaluation board is already populated with the necessary components for proper functioning of the ADC such as a differential amplifier, low-noise voltage reference, and low-jitter 16 MHz clock, and we can communicate with it using SPI.  The ADC uses a sigma-delta architecture and provides a number of different filters offering a trade-off between sampling rates and noise performance.  For our work, we have found that using the low-latency filter with a sampling rate of 31.25 kSPS provides optimal results with a 3 dB bandwidth of 13.7 kHz and a noise floor of 2.74 \si{\micro\volt} corresponding to $\delta I_p=411$ \si{\micro\ampere}.  The ADC has a voltage range of $\pm2.5$ V, corresponding to $\pm 375$ A with our choice of transducer and sense resistor; however, the range can be changed easily with a different sense resistor.  We store each measurement value in LPDDR memory which holds up to $4\times10^6$ values corresponding to over two minutes of continuous acquisition.  This data can be retrieved from memory using the PC link for later analysis.  In addition to the measurement used for feedback, we record an additional, out-of-loop ``validation'' signal $m(t)$ using an identical transducer, sense resistor, and ADC, and this signal is also stored in memory.

We realize our control law in programmable logic using a finite state machine that calculates a discrete approximation to the PID control law
\begin{equation}
u(t) = K_p e(t) + K_i\int_{-\infty}^t e(t')dt' + K_d\frac{de(t)}{dt}
\label{eq:PID}
\end{equation}
where $e(t)=r(t)-y(t)$ is the error signal and $K_p$, $K_i$, and $K_d$ are the proportional, integral, and derivative gain coefficients, respectively.  The control signal $r(t)$ is either a piece-wise sequence of linear ramps generated on-the-fly using a linear-ramp generator or is an arbitrary, pre-programmed value read from memory.  For each time step $n$ separated in time by $T_s$ (the inverse of the ADC sampling rate) we calculate the actuator value $u_n$ as\cite{Bechhoefer2005}
\begin{align}
u_n &= u_{n-1}+2^{-N}A\bigg[\tilde{K}_p(e_n-e_{n-1})+\frac{\tilde{K}_i}{2}(e_{n}+e_{n-1})\notag\\
&+\tilde{K}_d(e_n-2e_{n-1}+e_{n-2})\bigg]
\label{eq:ControlLaw}
\end{align}
where $A$ is an overall conversion factor needed to account for the different discretization steps between a voltage measured by the ADC and one output by the DAC, and each of the $\tilde{K}$ is the discrete equivalent of the PID gain coefficients in Eq.~\eqref{eq:PID}: $\tilde{K}_p=K_p$, $\tilde{K}_i=K_iT_s$, and $\tilde{K}_d=K_dT_s^{-1}$.  Each $\tilde{K}$ is represented as a 16-bit integer in programmable logic, and we approximate fractional gain values by shifting the sum of products right $N$ bits, implementing division by $2^N$.  We take care with the order of operations in Eq.~\eqref{eq:ControlLaw} as implemented in the FPGA so that values are not truncated prematurely and the full precision of the ADC is used.

We use the discrete control law in Eq.~\eqref{eq:ControlLaw}, which effectively calculates the correction to the actuator signal rather than the actuator signal itself, instead of the more direct discretization of Eq.~\eqref{eq:PID} for two main reasons.  The first is that the discrete integral term in Eq.~\eqref{eq:PID} is unbounded, and one needs to include additional logic to prevent integral wind-up when either input or output values saturate.  The second and more compelling reason is that our physical system is non-linear through the relationship between the MOSFET gate-source voltage and the drain-source conductance (inset of Fig.~\ref{fg:Diagram}), and we need robust, low-noise control of currents from 20 A to 400 A.  While we can approximate the dynamical relationship between $u(t)$ and $y(t)$ as linear over small intervals, this cannot be done over the entire range which crucially means that the optimum gain parameters for the controller change depending on the desired set-point $r(t)$.  Therefore, at each time step we update $u_n$ using not only the current and past values of the error signal but also according to pre-programmed values for each $\tilde{K}$ corresponding to different values of the set-point $r_n$.  Since Eq.~\eqref{eq:ControlLaw} calculates the actuator signal as the sum of the previous value and a correction, it is easily adapted to include time-varying gain parameters.  Equation~\eqref{eq:PID}, on the other hand, calculates the actuator anew at each time step, and changing the gain parameters between two different steps leads to significant jumps in the output signal due to the integral term.

\section{Tuning and modelling}
\label{sec:Tuning}
In many implementations of PID controllers in experimental physics laboratories the gain coefficients are manually tuned in an iterative process to achieve the best possible performance.  This technique is often used when either the performance simply needs to be ``good enough'' or when measuring the system's transfer function is not feasible.  With our system, we can drive the MOSFET gate-source voltage with sinusoidal signals of varying frequencies $\omega$, and, by changing the DC offset of these signals, we can measure the linearized AC system transfer function $G(\omega,I_p)$ about the steady-state DC primary current $I_p$ corresponding to a particular DC gate-source voltage.  Given a desired closed-loop response function $T(\omega)$, we calculate the appropriate control law $K(\omega,I_p)$ as\cite{Bechhoefer2005}
\begin{align}
K(\omega,I_p) &= \frac{T(\omega)}{1-T(\omega)}G^{-1}(\omega,I_p)M^{-1}(\omega)\notag\\
			  &= K_p(I_p) + \frac{K_i(I_p)}{i\omega} + i\omega K_d(I_p)\label{eq:LoopShaping}\\
G(\omega,I_p) &\approx \frac{G_0(I_p)}{1+\frac{i\omega}{\omega_1(I_p)}-\frac{\omega^2}{\omega_2^2(I_p)}},
\label{eq:OpenLoop}
\end{align}
where we approximate the dynamical system represented by $G(\omega,I_p)$ as second order with set-point dependent coefficients $G_0(I_p)$, $\omega_1(I_p)$, and $\omega_2(I_p)$.  For the low-latency filter that we use in our work, we can include an analytic representation of the ADC measurement response $M(\omega)$ as specified in the datasheet.  Given a desired first-order closed-loop response $T^{-1}(\omega) = 1+i\omega/\omega'$ with cut-off frequency $\omega'$ we determine the PID gain coefficients using
\begin{subequations}
\begin{align}
K_p(I_p) &= \frac{\omega'}{G_0(I_p)\omega_1(I_p)}\label{eq:PropGain}\\
K_i(I_p) &= \frac{\omega'}{G_0(I_p)}\label{eq:IntGain}\\
K_d(I_p) &= \frac{\omega'}{\omega_2^2(I_p)G_0(I_p)}\label{eq:DerivGain},
\end{align}
\label{eq:Gains}%
\end{subequations}
where we have neglected the measurement response to simplify the calculation of the coefficients.  This approximation is valid only when $\omega'\ll T_s^{-1}$.

\begin{figure}[tbp]
\includegraphics[width=\figwidth]{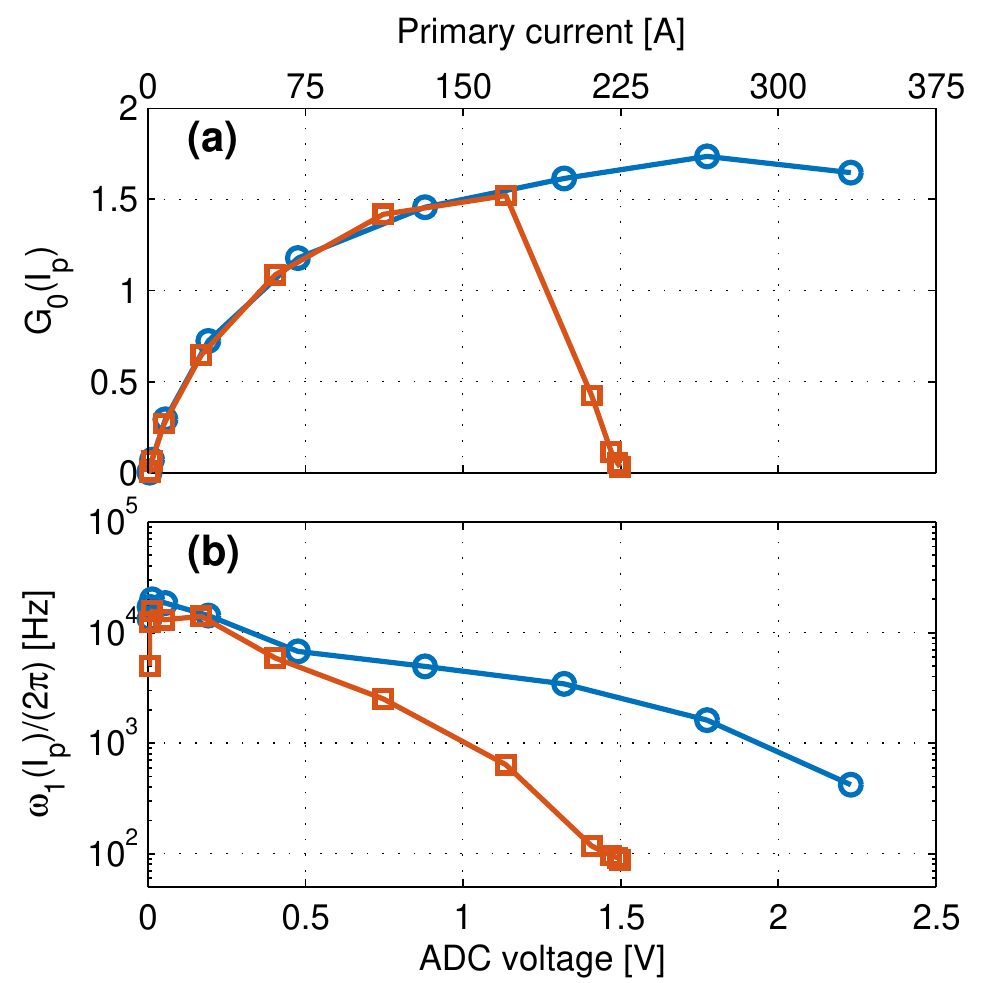}%
\caption{System response as a function of ADC voltage for power supply voltages of $12.5$ V (blue circles) and $7.2$ V (red squares).  (a) AC gain measured as the ratio of the voltage measured on the ADC and the driving voltage on the DAC.  (b) First-order frequency $\omega_1/(2\pi)$.}%
\label{fg:SystemResponse}%
\end{figure}

\begin{figure}[tbp]
	\includegraphics[width=\figwidth]{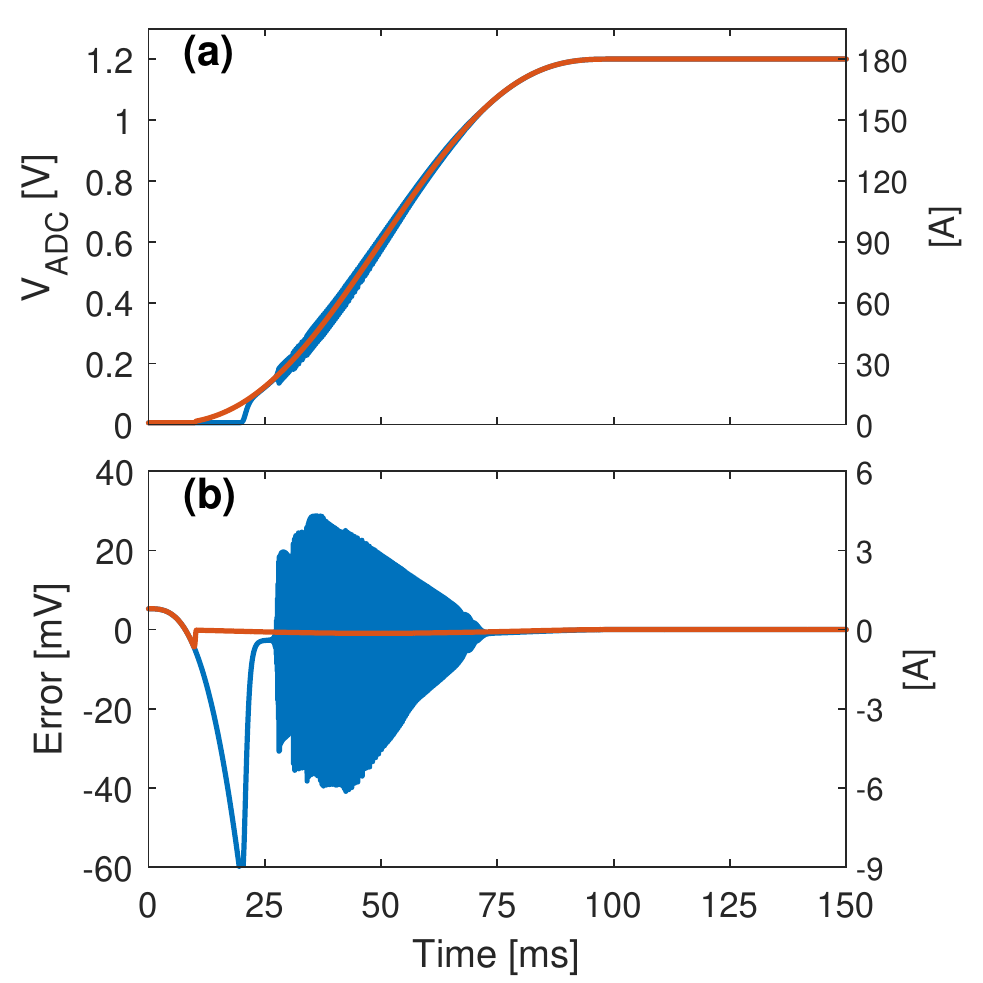}%
	\caption{Effect of varying the loop response with set-point at a power supply voltage of 7.2 V.  (a) Temporal profile of the measured voltage/current with (red) and without (blue) gain scheduling.  (b) Difference between measured voltage/current and set-point with (red) and without (blue) gain scheduling.}%
	\label{fg:LoopVariation}%
\end{figure}

Figures~\ref{fg:SystemResponse}a and b show parameters $G_0(I_p)$ and $\omega_1(I_p)$ as functions of the steady-state primary current $I_p$ for two different voltages 7.2 V and 12.5 V of the main power supply.  The second-order frequency $\omega_2$ is not shown because the additional term does not improve the model of the system response.
The lack of data beyond an ADC voltage of 1.5 V ($I_p = 225$ A) for a power supply voltage of 7.2 V is simply due to the total resistance of our system which limits the maximum current to 225 A.  For both power supply voltages we see a strong dependence of the system response on the desired primary current.  In particular, we see how the AC gain $G_0(I_p)$ for a power supply voltage of 7.2 V drops precipitously between 150 and 225 A as the MOSFETs' conductances saturate (see inset of Fig.~\ref{fg:Diagram}).  This gain maximum, present also at the higher power supply voltage of 12.5 V, presents a problem when one needs to sweep the primary current across that peak.  If the gain parameters are optimized for currents either before or after the gain peak then during the sweep the open-loop transfer function $K(\omega,I_p)G(\omega,I_p)$ will be larger than expected and the system may oscillate.  Alternatively, if the gain parameters are optimized for the gain peak, then the controller will perform sub-optimally at currents where the AC gain is lower than its peak value.  By using gain scheduling -- namely, adjusting the gain parameters with the desired primary current -- we can ensure that we have an optimum loop response during the entire sweep.  Figure~\ref{fg:LoopVariation} compares the effect of varying the loop response as a function of the set-point with a conventional fixed response.  For this measurement, we have set the closed-loop response to behave as a low-pass filter with a cut-off frequency of 2 kHz (see below).
As can be seen in Figs.~\ref{fg:LoopVariation}a and b, the excessive gain as the current increases in the fixed response mode causes the system to oscillate.  Additionally, the lack of gain at the beginning of the current rise leads to a slow turn-on as the gate-source voltage slowly accumulates due to the integral term.  Using gain scheduling eliminates oscillations during the initial ramp from 0 to 1.2 V and improves the initial turn-on behaviour.

\begin{figure}[tbp]
\includegraphics[width=\figwidth]{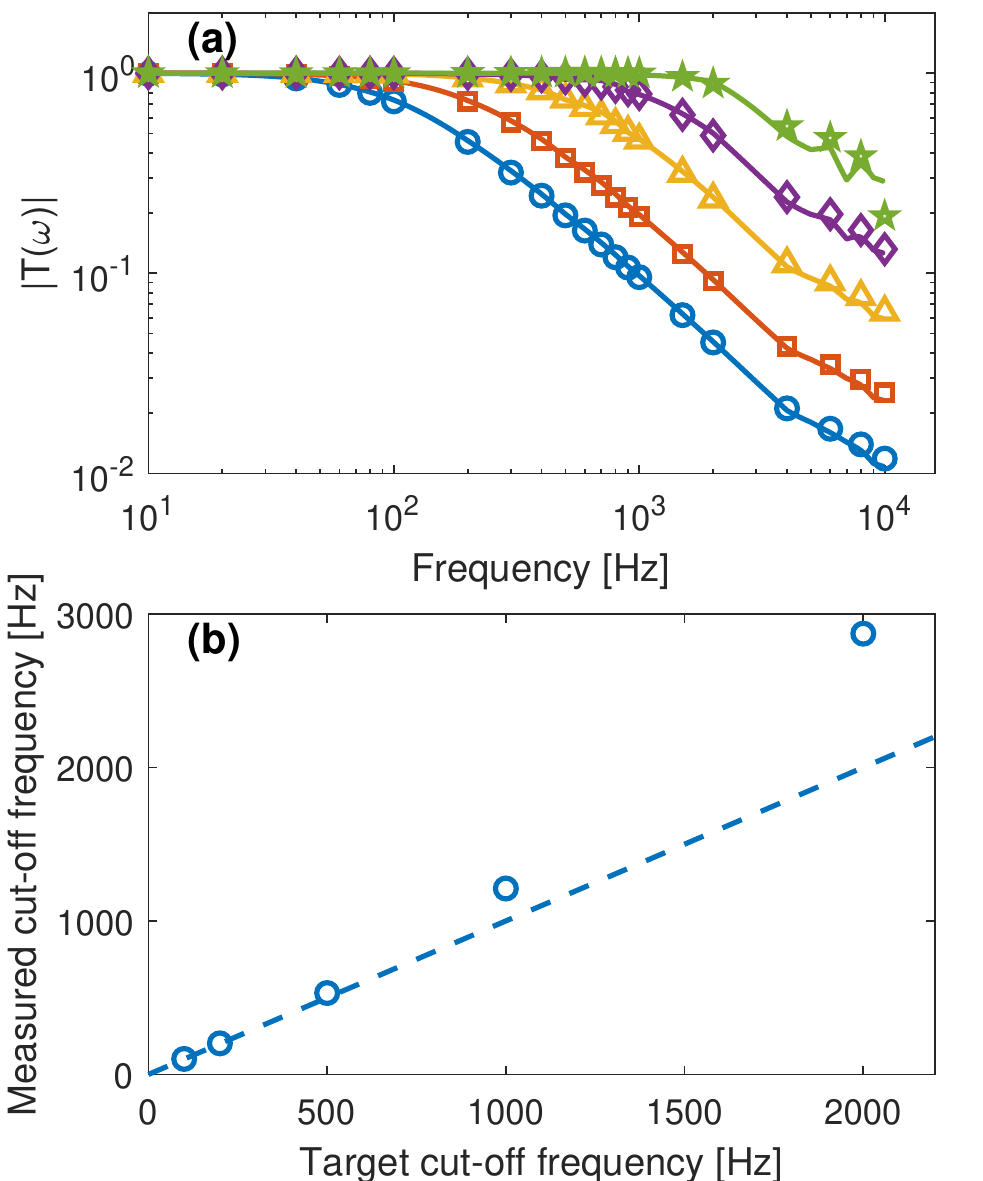}%
\caption{Closed-loop response as a function of frequency for different desired low-pass filter frequencies with a power supply voltage of 12.5 V.  (a) Measured amplitude of the closed-loop response $|T(\omega)|$ at a set-point of $V_{\rm ADC} = 1$ V for target frequencies of 100 Hz (circles), 200 Hz (squares), 500 Hz (triangles), 1 kHz (diamonds), and 2 kHz (stars).  Solid lines are the prediction from the measured open-loop response and a model of the controller's behaviour.  (b) Measured 3 dB cut-off frequency compared to the target cut-off frequency.  Circles are the measurements, and the dashed line gives the ideal behaviour.}%
\label{fg:ClosedLoopResponse}%
\end{figure}

Using the measured values of $G_0(I_p)$ and $\omega_1(I_p)$ in combination with Eq.~\eqref{eq:Gains} we program the controller to have a low-pass filter closed-loop response with various cut-off frequencies $\omega'$; the results are shown in Figs.~\ref{fg:ClosedLoopResponse}a and b.
To measure the closed-loop response we increase the current in the Helmholtz coils from 0 A to 150 A ($V_{\rm ADC} = 1$ V) in 100 ms using a minimum-jerk trajectory\cite{Chisholm2018} to minimize transients before modulating the control signal with a sinusoidal signal of varying frequency and fixed amplitude of 20 mV which corresponds to a primary current amplitude of 3 A.  We model the response by multiplying our measured $G(\omega,I_p)$ with $M(\omega)$ and the Fourier transform of Eq.~\eqref{eq:ControlLaw}.  It is important to include both the latency of the ADC, equal to $T_s=32$ $\mu$s, and the latency of the FPGA's PID process which comprises the time to read from the ADC ($2.6$ $\mu$s), the time to process the data (11 clock cycles, or $0.22$ $\mu$s), the time to write to the DAC ($2.6$ $\mu$s), and the time for the DAC output to change ($\approx1$ $\mu$s) for a total latency of $\approx 6$ $\mu$s.  Figure~\ref{fg:ClosedLoopResponse}a shows that we get excellent agreement between our measured response and the response expected from our model, with some deviations at high target cut-off frequencies.  The actual cut-off frequencies, shown in Fig.~\ref{fg:ClosedLoopResponse}b, differ significantly from the expected values for high cut-off frequencies, and this is a consequence of modelling the physical system as only a second-order dynamical system and neglecting the measurement response when calculating the gain coefficients.  A more complete description would result in better correspondence between the desired and measured cut-off frequencies but would also require a more complex control law.

\section{Performance}
\label{sec:Performance}
\begin{figure}[tbp]
	\includegraphics[width=\figwidth]{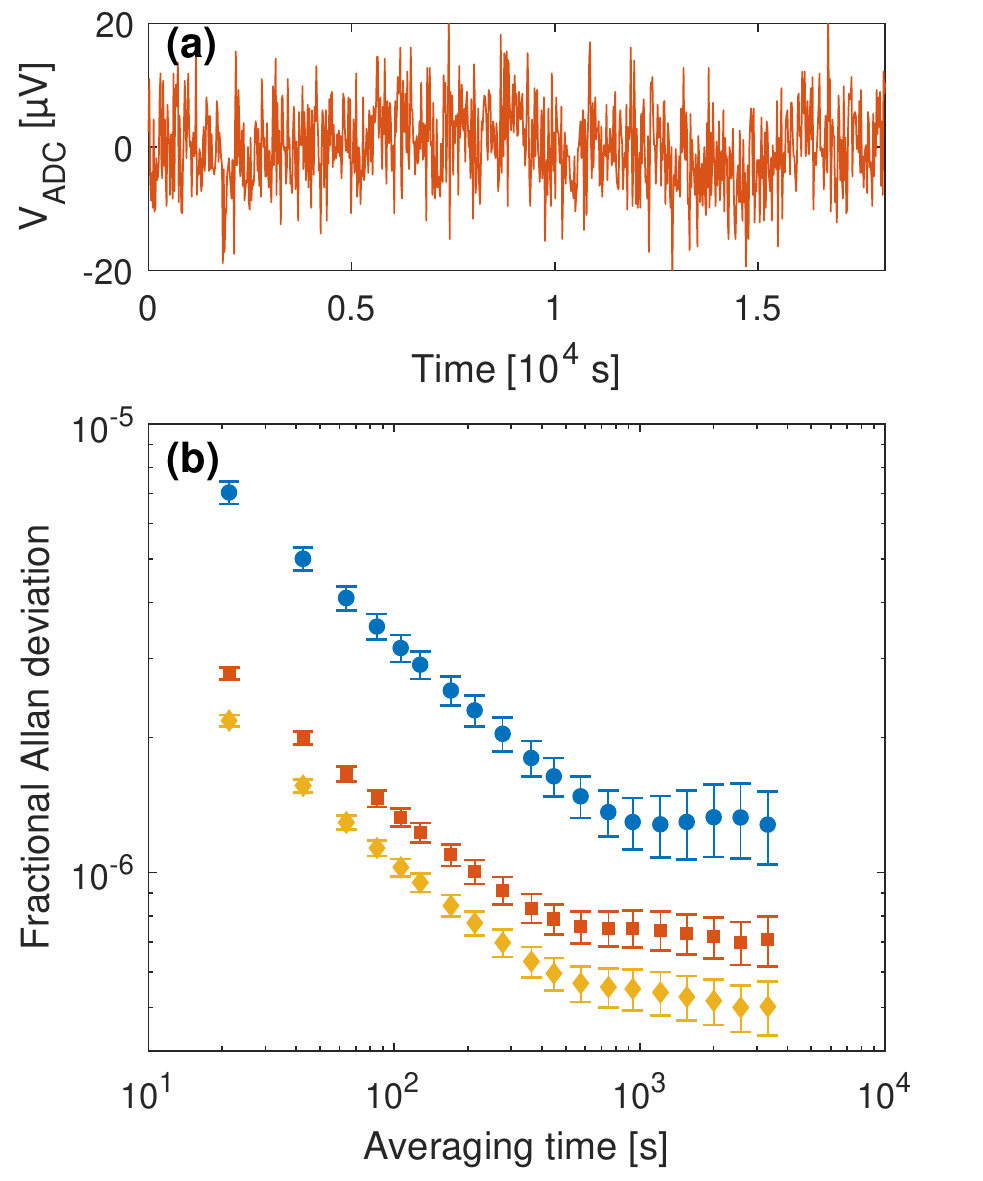}%
	\caption{Measuring the stability of the steady-state current across experimental cycles with a power supply voltage of 12.5 V.  (a) Time trace of the validation ADC for a set-point of 337.5 A sampled at an offset time of 500 ms from the start of the ramp with which we calculate an estimate of the Allan deviation.  The mean of the trace has been subtracted for clarity.  (b) The fractional Allan deviation for set-points of 150 A (blue circles) and 337.5 A (red squares) for offset times between 250 ms and 950 ms.  The yellow diamonds are an estimate of the limiting fractional Allan deviation at a set-point of 337.5 A (see text).  Error bars indicate 1-$\sigma$ confidence interval on the estimate of the fractional Allan deviation.}%
	\label{fg:AllanDeviation}%
\end{figure}
For our particular experiment, Feshbach resonance spectroscopy, we are primarily concerned with how much variability and noise there is in the current regulated by our controller when the set-point is fixed, and hence the variability in the magnetic field experienced by the atomic sample inside the Helmholtz coils.  Both fluctuations in the current during an experimental cycle (AC) and between cycles (DC) are of interest to us; the former can ``wash out'' features of interest and lead to short coherence times for superpositions of magnetically sensitive states, while the latter compromises the reproducibility of any measurement and ultimately limits its accuracy and precision.  We quantify the DC variability using the fractional Allan deviation, shown in Fig.~\ref{fg:AllanDeviation}, which is the Allan deviation of the validation ADC voltage divided by mean voltage over many realisations.  To measure the Allan deviation we increase the current from 0 A to the desired set-point (either 150 A or 337.5 A) in 100 ms using a minimum-jerk ramp to reduce transients, and then hold the set-point steady for 900 ms before turning the current off.  The loop parameters are set to give a nominal closed-loop response of a 2 kHz low-pass filter with the gain coefficients chosen to optimize the performance at each set-point individually.  We measure the time-dependent voltage for 852 realizations with a repetition period of approximately 23 s, corresponding to five hours worth of data for each of the two set-points.  Both the voltage from the feedback ADC and the validation ADC are recorded.  To estimate the Allan deviation, we pick a particular offset time after the minimum-jerk ramp has finished and then calculate the Allan deviation using the 852 independent realizations of the process (see Appendix); an example of the validation signal for which the Allan deviation is calculated for an offset time of 500 ms after the start of the minimum-jerk ramp is shown in Fig.~\ref{fg:AllanDeviation}a for a set-point of 337.5 A.  We then estimate the distribution of fractional Allan deviation by calculating the Allan deviation for 1000 equidistant offset times from 250 ms to 950 ms after the ramp has started, and we plot the fractional Allan deviations with their 1-$\sigma$ confidence intervals in Fig.~\ref{fg:AllanDeviation}b.

For the 150 A set-point the fractional Allan deviation reduces to 1 ppm at an averaging time of 900 s while for the 337.5 A case the fractional Allan deviation attains $<1$ ppm at 550 s before $1/f$ noise dominates and the Allan deviation reaches a minimum.  We estimate the stability limit for a set-point of 337.5 A by calculating the Allan deviation of the difference between the feedback and validation measurements, which should remove common-mode fluctuations, and dividing the result by $\sqrt{2}$ as the variance of the difference of two uncorrelated signals is the sum of the variances of the signals.  We infer that our feedback scheme has an Allan deviation that is only about 30\% higher than the limit we calculate.  Subsequent investigation of the two measurement chains for feedback and validation has suggested that the plateau reached in the Allan deviation near $10^3$ s is due to fluctuations in the offset secondary current of the feedback transducer which is not present on the validation transducer.  It is expected that replacing the feedback transducer will lead to even better stability which will then be limited by thermal variability in the components of the measurement chain.

\begin{figure}[tbp]
	\includegraphics[width=\figwidth]{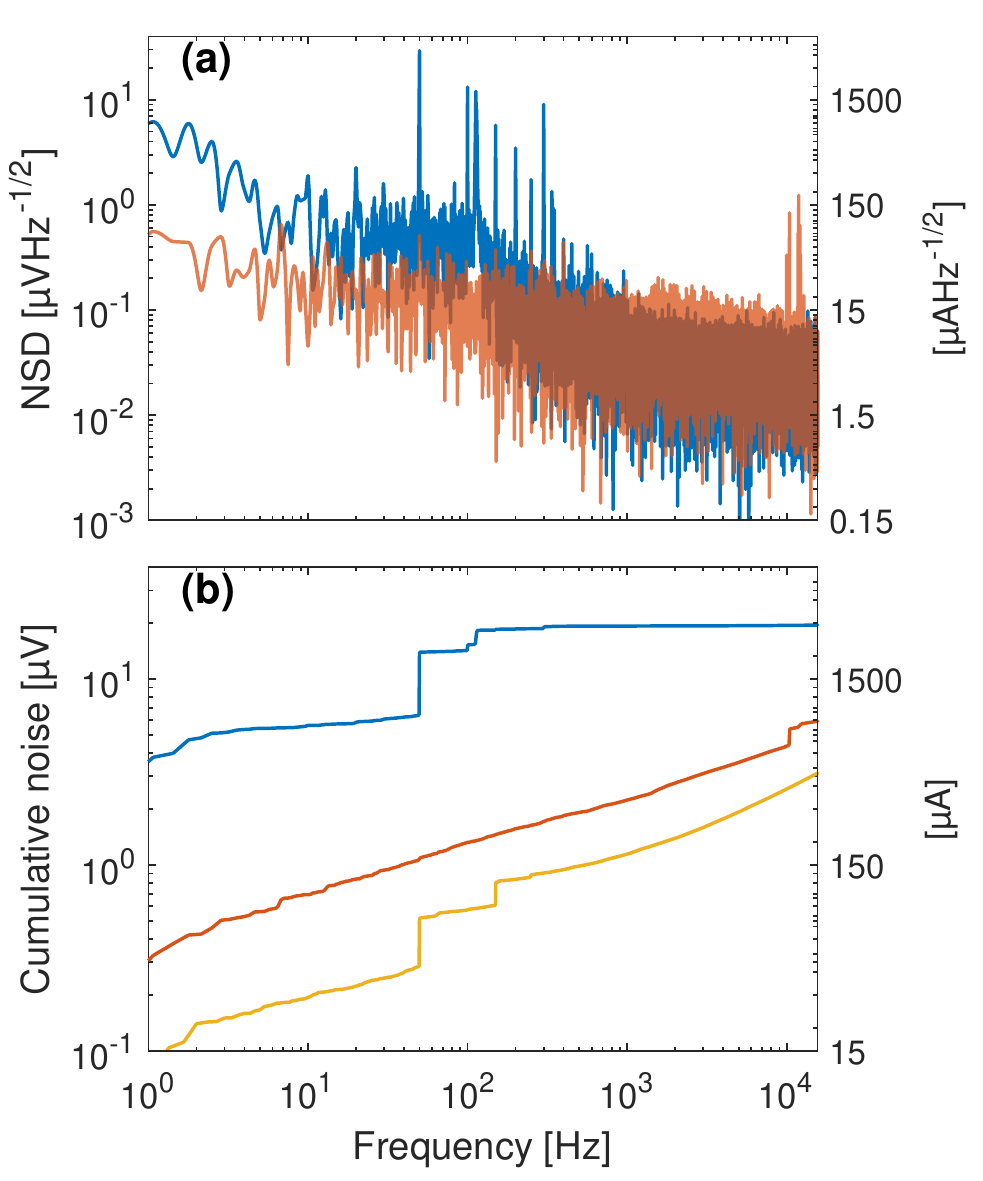}%
	\caption{Frequency-domain analysis of the validation signal for a power supply voltage of 12.5 V and for a steady-state set-point of $V_{\rm ADC} = 1$ V ($I_p = 150$ A).  \chk(a) Noise spectral density (NSD) of the error signal for a power supply voltage of $12.5$ V.  Red curve is with our servo engaged, blue curve is using the power supply in current-controlled mode.  (b) Cumulative RMS noise in the validation signal.  Colours mean the same as in (a), and the yellow curve is the baseline noise with no current.  Axes on the right give the equivalent quantities when converted to current.}%
	\label{fg:Spectral}%
\end{figure}

In Fig.~\ref{fg:Spectral} we show the spectral characteristics of the steady state validation signal using the same ramp as for measuring the Allan deviation but extending the holding period at a set-point of $V_{\rm ADC} = 1$ V to 5 s.
Fig.~\ref{fg:Spectral}a displays the noise spectral density (NSD) of the out-of-loop ADC voltage both when the servo is engaged using optimal parameters and when we turn off the servo and use the power supply in its native current-controlled mode.  Since the power supply's internal servo is much slower than our controller, with a bandwidth of $\sim1$ Hz, we wait 1 s until the current has reached a steady-state before calculating the NSD.  Noise in the error signal with our external servo engaged is strongly suppressed below the nominal 2 kHz control bandwidth compared to the power supply's internal servo.  Above 2 kHz the NSD of our external servo exceeds that of the internal servo, in large part due to the $T_s = 32$ $\mu$s latency of our ADC.  When $\omega T_s\ll 1$ the phase shift due to the filter latency does not significantly contribute to the overall phase shift of the system; however, when $\omega T_s \sim 1$ -- which occurs at $\omega/(2\pi) \sim 5$ kHz -- the latency of the filter limits the performance of the controller as a whole.  For our system, our model predicts that at frequencies near 5 kHz our controller amplifies noise rather than suppressing it.  This could potentially be improved with a higher sampling rate at a cost of more overall noise in the control loop.  Noise peaks at $\approx 10$ kHz and 12 kHz are picked up from external sources as a result of imperfect grounding between the DAC and the power supply.  By looking at the cumulative noise (Fig.~\ref{fg:Spectral}b) we see that our servo operates with a factor of 3 to 10 times less AC noise than the power supply alone, and that it is only a factor of two above the noise floor of our measuring system.  It is worth emphasizing that this measurement neglects fluctuations in the DC level of the current for which our servo has a factor of one hundred better stability.

\begin{figure}[tbp]
\includegraphics[width=\figwidth]{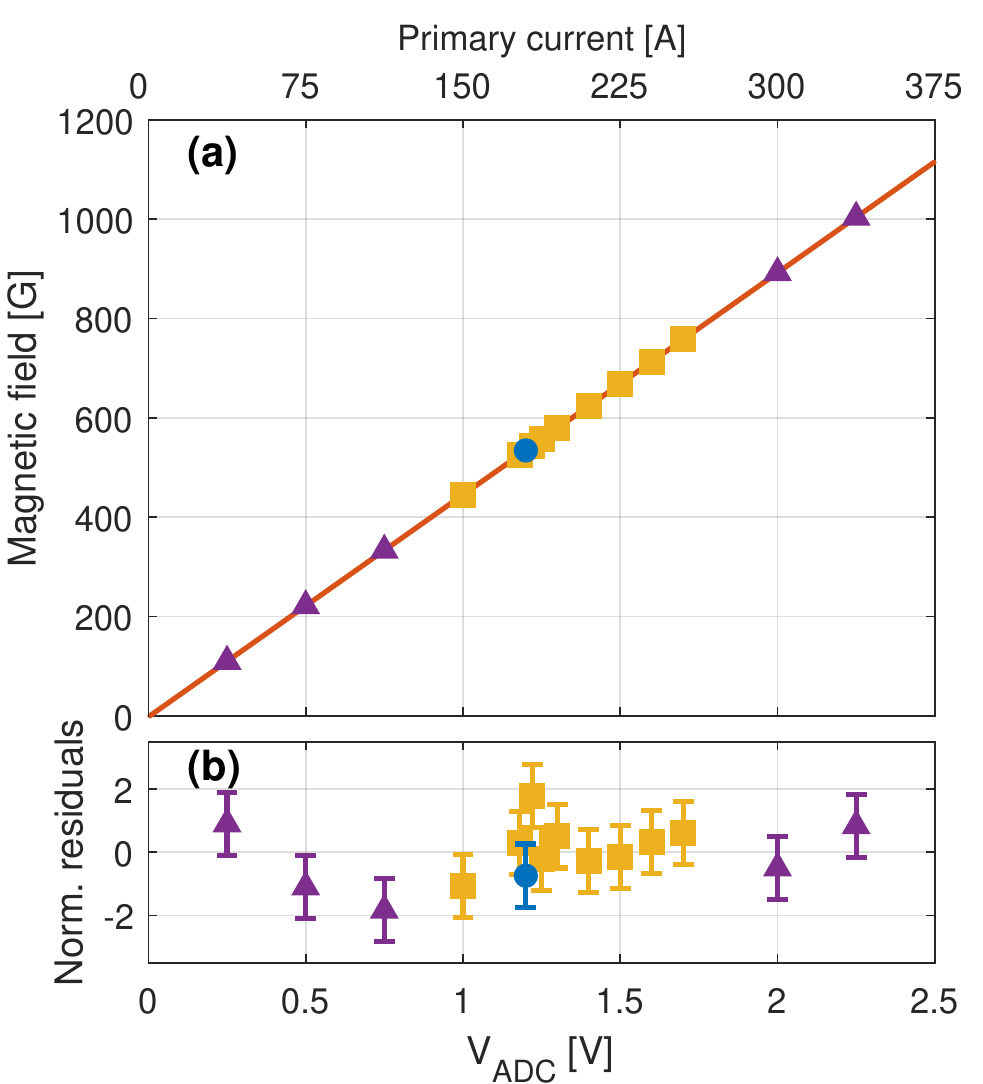}%
\caption{Measurement of the magnetic field produced by our system as a function of voltage set-point using Rabi spectroscopy of an atomic sample.  Shape and colour of the markers indicates the day on which the measurement was made: day 0 (blue circles), day 1 (yellow squares), and day 4 (purple triangles).  (a) Measured magnetic fields (markers) and linear fit (red line).  Error bars are smaller than the width of the line and are not shown.  (b) Residuals of the fit normalized to individual measurement uncertainties.}%
\label{fg:Calibration}%
\end{figure}

Finally, we demonstrate that our system is capable of generating the required magnetic fields for Feshbach resonance studies.  To measure the magnetic field, we first prepare a sample of $^{87}$Rb atoms in a crossed-beam optical dipole trap and in the $\left|F=2,m_F=0\right\rangle$ state at a low field (9 G) using a separate, low-current power supply.  We then turn off the low-current supply and use our high-current system to increase the current from 0 A to a particular set-point in 75 ms using a minimum-jerk trajectory.  We then wait for 50 ms for any transients to decay.  At this point, we use Rabi spectroscopy of the $\left|F=2,m_F=0\right\rangle\rightarrow\left|F=1,m_F=-1\right\rangle$ transition with a Rabi frequency of $\Omega/(2\pi) \approx 3$ kHz and a pulse time of 50 $\mu$s to determine the transition frequency from which we calculate the magnetic field.  Due to the long duty cycle of our experiment ($\sim100$ s per cycle), these data were taken over the span of five days.  The results are shown in Fig.~\ref{fg:Calibration} with the day the measurement was taken indicated by the shape and colour of the marker.
We see that we get a highly linear relationship between the set-point and the measured magnetic field over an order of magnitude in magnetic field, and we can predict the magnetic field for other set-points with a maximum uncertainty of 5 mG, making our system suitable for studying Feshbach resonances in a range of atomic systems.  The constant of proportionality between magnetic field and current that we measure is consistent with previous measurements that we have made using an ultra-stable commercial power supply at lower ($<10$ A) currents.  By using a linear regression on the fit residuals as a function of the day on which the data were taken, we estimate the magnetic field drift at $-3\pm25$ mG/month which is consistent with past magnetic field measurements that displayed an approximate $\pm20$ mG/month drift in the magnetic field experienced by the atoms.  This drift is likely due to changes in the cooling water temperature for the Helmholtz coils and changes in the background magnetic field rather than changes in the resistor's load value ($<50$ ppm over 2000 h of use), the ADC voltage reference ($<20$ ppm over 2000 h or use), or the transducer's offset current ($<0.1$ ppm/month).

\section{Conclusion}
\label{sec:Conclusion}
In this article we have presented a digital PID controller for stabilizing the large currents needed to generate magnetic fields for studies involving Feshbach resonances.  The performance of our controller is easily tailored through a digitally programmable loop response which can implement gain scheduling, allowing for robust and precise control of non-linear systems.  We achieve a fractional stability of $<1$ ppm at 337.5 A in a 10 minute averaging time and with a 2 kHz control bandwidth.

We have already used our PID controller to study the behaviour of a Feshbach resonance in the collision of $^{87}$Rb and $^{40}$K atoms at energies far above threshold\cite{Thomas2018}.  Our system could be used for studying narrow Feshbach resonances at large magnetic fields, such as higher angular momentum Feshbach resonances or resonances in non-alkali metal systems\cite{Barbe2018}, and it can also find use in other cold atom experiments such as ion trapping\cite{Merkel2019}.

\section*{Acknowledgements}
This work was funded by the New Zealand Tertiary Education Commission through the Dodd-Walls Centre for Photonic and Quantum Technologies.

\section*{Supplementary Material}
Hardware definition language (HDL) code for the FPGA architecture as well as MATLAB\textsuperscript{\textregistered} software (including documentation) for controlling the FPGA from a computer can be found on GitHub\cite{GitHub}.

\section*{Appendix: Calculating the Allan deviation}
To calculate the Allan deviation for the steady state current signal we perform 852 independent measurements comprising an initial 100 ms minimum-jerk ramp from 0 A to the set-point followed by 900 ms of hold followed by the current being switched off and the coils being allowed to cool down for about 22 s.  This process gives us a set of measurements $I_j(t_k)$ where $j$ indexes the measurement cycle ($j\in [0,851]$) and the $t_k$ are the times since the start of the ramp with $0 \leq t_k < 1$ s and $t_{k+1}-t_k = 32$ \si{\micro\second}.  For a fixed $t_k$, $I_{j+1}(t_k)$ is measured 23 s after $I_j(t_k)$.  For 1000 equally-spaced $t_n \in [\SI{250}{\milli\second},\SI{950}{\milli\second}]$, we calculate the Allan variance
\[
\sigma_n^2(\tau) = \frac{1}{2}\left\langle \left(\bar{I}_{j+1}(t_n) - \bar{I}_{j}(t_n) \right)^2\right\rangle
\]
where $\bar{I}_{j}(t_n)$ is the $j$th average over an interval of time $\tau=m\tau_0$ with $m$ an integer and $\tau_0=\SI{23}{\second}$ for fixed $t_n$.  In practice, we use the overlapped estimator of the Allan variance\cite{IEEEStandard} which is calculated as
\begin{align}
\sigma_n^2(m\tau_0) &= \frac{1}{2(N-2m)m^2\tau_0^2}\sum_{k=0}^{N-2m-1}\left[ \sum_{j=m+k}^{2m+k-1}I_j(t_n)\right. \notag\\
&\left.- \sum_{j=k}^{m+k-1}I_j(t_n) \right]^2\notag.
\end{align}
The Allan deviation is then $\sigma_n(m\tau_0) = \sqrt{\sigma_n^2(m\tau_0)}$.  From the set of independent estimates of the Allan deviation at a particular averaging time $\tau$, we compute the mean and variance of the Allan deviation over the set of estimates indexed by $n$.

\end{document}